\magnification\magstep1

\input mssymb
\font\twelvebf=cmbx12
\font\eighteenbf=cmbx18
\font\ninerm=cmr9
\font\ninebf=cmbx9

\def\abs#1{{\left\vert{#1}\right\vert}}
\def\bigo#1{{O\bigl(#1\bigr)}}
\def\sorted#1{{\hbox{\sl sorted}\left(#1\right)}}

\vskip 24pt
\centerline{\eighteenbf Finding Matches between Two Databases}
\centerline{\eighteenbf on a Quantum Computer}
\vskip 18pt
\vbox{
Mark Heiligman 
\sl
\vskip .5pt
National Security Agency, Suite 6111
\vskip .5pt
Fort George G. Meade, Maryland 20755
\vskip .5pt
\rm
mheilig@zombie.ncsc.mil
}

\vskip 18pt
\vbox{\noindent{\ninebf Abstract:} \ninerm
Given two unsorted lists each of length N that have a single common
entry, a quantum computer can find that matching element with a work
factor of $\displaystyle O(N^{3/4}\log N)$ (measured in quantum memory
accesses and accesses to each list).  The amount of quantum memory
required is $\displaystyle O(N^{1/2})$.  The quantum algorithm that
accomplishes this consists of an inner Grover search combined with a
partial sort all sitting inside of an outer Grover search. }

\everypar={} 
\def\preskip{15pt}
\def\postskip{8pt}
\def\bskip{6pt}

\vskip \preskip
\vskip \preskip
\noindent{\twelvebf Introduction}
\vskip \postskip

Given two databases, each containing $N$ entries, an important
question is often to locate a common entry in both databases, if one
exists. If the databases are unstructured, then this can be a hard
problem. This note gives a quantum mechanical algorithm for solving
this matching problem in time $\bigo{N^{3/4}\log N}$, which is better
than the best known classical algorithm for this problem.

Although this bears some superficial resemblance to the collision
problem, for which a quantum mechanical solution was given in [BHT],
the question being addressed here actually is quite different.  In the
collision problem, stated as a database problem, it was assumed a
single database contained every entry exactly twice, and the goal was
to find a single example of a duplicate entry.  Here, there are two
databases and it is assumed that there is exactly one entry shared by
both of them, and the problem is to locate that one entry in both
databases. 

The solution here, in common with the solution to the collision
problem given in [BHT], uses Grover's search algorithm as a
subroutine, and all of the quantum mechanics is relegated to what goes
on inside of Grover's algorithm.  All that needs to be understood for
using Grover's algorithm is what its resource requirements are.
Another common feature of the solution here with the collision problem
is the use of a sorting subroutine on a sublist and the fact that a desired
item can be located quickly in a sorted list, a procedure often
referred to as ``insertion into a sorted list''.

The cost of running an algorithm is equivalent to the time it takes to
run.  In the model here, static memory does not add to the cost, but
reads and writes from and to memory do count.  In addition, access to
the databases also count towards the cost.  Since all cost estimates in
this note are done up to a constant factor, all that really needs to
be counted are the number of internal memory accesses for the
algorithm and the number of databases queries that the algorithm makes.

\vskip \preskip
\noindent{\twelvebf The Classical Solution to the Matching Problem}
\vskip \postskip

It may be useful to begin with an exact statement of the problem to be
solved:  There are two lists $L_1$ and $L_2$ each of length $N$, and
there is a element $l_0$ common to both $L_1$ and $L_2$.  The object
is to find that common element.

An exhaustive solution to the matching problem would be simply to
check all possible pairs, where the first element of each pair comes
from $L_1$ and the second element of each pair comes from $L_2$.
Since there are $N^2$ such pairs, the total amount of work is
$\bigo{N^2}$, where the implied constant contains the work of
accessing the pair of elements from the two lists and the work in
comparing the two elements to see if they match.

There is a much better classical solution, however.  Sort the first
list $L_1$ and then run down all the elements of $L_2$ and check to
see which one occurs in $\sorted{L_1}$.  The cost of sorting $L_1$
using an efficient classical sorting algorithm is $\bigo{N\log N}$.
Such sorting algorithms are discussed in detail in [K], [NR] and [S],
and can even be done in place.  The constant in the notation here
includes the cost of accessing two elements from $L_1$, the cost of
comparing these two elements, and the cost of swapping these two
elements.  To check a single element of $L_2$ against $\sorted{L_1}$
is the problem of insertion into a sorted list, and using a standard
divide-and-conquer strategy, can be accomplished with cost $\bigo{\log
N}$, where the constant here absorbs the cost of accessing an element
from $\sorted{L_1}$ and the cost of comparing this element to the
element of $L_2$ being tested.  Since there are $N$ elements of $L_2$,
the entire cost of searching through all of $L_2$ is $\bigo{N\log N}$,
and the entire cost of the algorithm is therefore $\bigo{N\log N}$.

Of course, it is also possible to sort $L_2$ instead and then search
through $L_1$ for the match.  Alternatively, it is possible to sort
both $L_1$ and $L_2$ producing two sorted lists, $\sorted{L_1}$ and
$\sorted{L_2}$ with cost $\bigo{N\log N}$ and then go down both sorted
lists with a pair of pointers looking for a match.  The cost of this
last step is $\bigo{N}$, so the total cost is still $\bigo{N\log N}$.

\vskip \preskip
\noindent{\twelvebf Grover's Algorithm for Quantum Searching}
\vskip \postskip

In a very interesting and widely referenced paper [Gr], Lov Grover has
shown that searching an unordered list can be done much faster on a
quantum computer than on a classical computer.   Classically for a
list of size $N$, searching for a single item will cost $O(N)$
accesses to the list on average.  Grover's quantum algorithm only
costs $O\bigl(\sqrt{N}\bigr)$ accesses to the list.  A good analysis
of Grover's algorithm can be found in [BBHT].  It has been shown in
[Z] that with no other structure present in the list, the best than
can be done with a quantum algorithm is $O\bigl(\sqrt{N}\bigr)$.

Use of the Grover search algorithm as a subroutine is what really
constitutes the quantum part of the new algorithm.  It isn't
absolutely necessary to go into the specifics of how Grover's
algorithm works in detail to understand how it is used, but it is
important to state what Grover's algorithm actually does and the cost
associated with invoking it as a subroutine.

Abstractly, Grover's algorithm goes as follows: Start with some
arbitrary function $F:\,X\rightarrow Y$ between finite sets having no
{\it a priori} structure.  Given some $y_0\in Y$, the object is to
find an $x\in X$ with $F(x)=y_0$, provided such an $x$ exists.  If
$k=\abs{\{x\in X \mid F(x)=y_0\}}$ denotes the number of different
solutions, then Grover's algorithm gives finds such a solution
(provided that one exists) after an expected number of
$O\bigl(\sqrt{\abs{X}/k}\bigr)$ evaluations of the function $F$ (which
must be implemented quantum mechanically).  Although this doesn't give
an exponential speedup, it is considerably better than exhaustively
searching through all of $X$.  In the present application, $k=1$
since there will be (at most) one possible solution.  In the usual
formulation of Grover's algorithm $\abs{X}$ is a power of 2, and for
the application here, this is not a serious restriction.

The usual way that Grover's algorithm is invoked is to compute an
indicator function $\chi_{F,y_0}$ defined by
$\chi_{F,y_0}(y_0)=1$ if $F(x)\ne y_0$ and $\chi_{F,y_0}(x)=0$ for all
$x$ with $F(x)\ne y_0$.
In order to compute $\chi_{F,y_0}(x)$, the the value of $F(x)$ will
need to be computed, and this has to be done quantum mechanically.
Although $F(x)$ may be described classically, it is not necessary that
$F(x)$ be computed classically through reversible classical logic, and
if there is a quantum algorithm that computes $F(x)$ more efficiently,
then it is perfectly acceptable to call that algorithm as a subroutine
from Grover's algorithm.

In evaluating the cost of Grover's algorithm, both the size of the
search space and the cost of evaluating $F$ enter.  Often the
discussion of Grover's algorithm ignores the cost of computing $\chi_{F,y_0}$
$O\bigl(\sqrt{\abs{X}/k}\bigr)$ times, but as will be seen, this cost
really can enter into the analysis in a significant way, and this is
because $F$ itself can be a very complicated function that may in fact
embody various search and sort algorithms.  Thus in the case that there
is at most one solution $x$ to $F(x)=y_0$, the cost of finding this
solution (and checking that it is a solution) is
$O\bigl(\sqrt{\abs{X}}\,C_F\bigr)$ where $C_F$ is the cost of a single
evaluation of $F$ (or equivalently, the cost of a single evaluation of
$\chi_{F,y_0}$). 

One other factor that needs to be mentioned is the success rate of
Grover's algorithm.  If there is at most one solution to $F(x)=y_0$,
then Grover's algorithm will find that solution with probability
$1-\epsilon$ where $\epsilon=\bigo{\abs{X}^{-1}}$.

It is important to note that applying Grover's algorithm in a naive
way to the problem of finding a matching pair of elements in two lists
each of size $N$ doesn't really yield much of a win.  If the lists are
both of size $N$, then checking all $N^2$ possible pairs of elements
for a match with Grover's algorithm would require $\bigo{\sqrt{N^2}} =
\bigo{N}$ steps. This should be compared to the classical algorithm
described above, which has an overall cost of $\bigo{N\log N}$.  Other
than in improvement by a logarithmic factor, the only aspect in which
Grover's algorithm really improves on the classical algorithm is in
memory usage.  The classical algorithm require a memory size of
$\bigo{N}$ to hold the sorted list (and to actually do the
sorting), while Grover's algorithm doesn't really use any memory
(other than what is used in computing single elements on each of the
lists and comparing them).  Thus from a pure efficiency standpoint, a
naive application of Grover's algorithm to the matching problem is not
really better than what a classical algorithm would do.

\vskip \preskip
\noindent{\twelvebf Quantum Matching}
\vskip \postskip

If the first list $L_1$ is sorted, and the Grover search only goes
over the second list, the cost of sorting $L_1$ is $\bigo{N\log N}$,
and there will be no improvement over the classical solution.  On the
other hand, notice that Grover search over the second list costs
$\bigo{\sqrt{N}}$ insertions into $\sorted{L_1}$, each at a cost of
$\bigo{\log N}$, and therefore the total cost of Grover searching over
the second list to find a match in $\sorted{L_1}$ is
$\bigo{\sqrt{N}\log N}$, which is dominated by the cost of the sort of
$L_1$.  If it were possible to only do the sort on a set of size
$\sqrt{N}$ instead of a set of size $N$, then the cost of the sort
would be $\bigo{\sqrt{N}\log N}$ also.

This leads to the main idea.  Suppose that the list $L_1$ is broken
into blocks of size $\sqrt{N}$, there being $\sqrt{N}$ such blocks.
If the block containing the match could be identified, then sorting
this block and Grover searching over $L_2$ would find the matching
element in time $\bigo{\sqrt{N}\log N}$.  The idea then is to identify
the block containing the match with an outer application of Grover's
algorithm as well.  The inner algorithm just described is easily
modified to return a simple yes/no answer on whether a block contains
a match to an element in $L_2$.  Using this as an indicator function,
the cost of running the outer Grover algorithm on the set of $\sqrt{N}$
blocks is $\bigo{\root 4 \of N}$ times the cost of computing this
indicator function.  The overall cost is therefore $\bigo{N^{3/4}\log
N}$. 

In this outer loop, the indicator function is computed quantum
mechanically (and in fact in a superposition of basis states), which
implies doing the sort and inner search quantum mechanically, and then
also undoing them once the yes/no answer has been computed, and in
fact, this whole procedure is repeated $\bigo{\root 4 \of N}$ times.
This may seem a little bizarre, because the actual sort algorithm is
now being performed on a superposition of lists, each list now
comprising a single block of size $\sqrt{N}$, and of course, the
outcome of the comparison and conditional swaps occurring in each sort
operation is also being done in superposition.  Nevertheless, however
classically nonsensical this is, the laws of quantum mechanics do not
seem to forbid such activities.

\vskip 24pt
\noindent{\twelvebf Error Analysis}
\vskip \preskip

As originally conceived, Grover's algorithm is probabilistic in
nature, with the correct result of a database search being produced
with a probability very nearly equal to 1.  However, in this
Grover-within-Grover algorithm being proposed here, a more refined
error analysis may be called for.

Each inner application of the Grover algorithm is a search on $L_2$
which is a list of length $N$, so the probability of any single inner
Grover subroutine call succeeding is $1-N^{-1/2}$.  The outer Grover
algorithm calls this subroutine $\bigo{N^{1/4}}$ times, so the
probability of all these calls returning the correct value for the
indicator function is $\bigl(1-N^{-1}\bigr)^{N^{1/4}}$ which is
still very close to 1 for large $N$.  Another way of seeing this is to
consider the expected number of failures of the inner Grover
subroutine call.  Since the probability of failure of any single call
is about $\bigo{N^{-1}}$ and this routine is call $\bigo{N^{1/4}}$
times, the expected number of failures is $\bigo{N^{-3/4}}$ which is
well below 1 for $N$ large.

In any case, single isolated failures of the inner Grover subroutine
call are not going to be disastrous for the whole algorithm, since the
indicator function for the correct block would simply return 0 on all
its arguments, which in turn would effectively mean that the outer
Grover iteration would be run one less time than optimal.  The result
is that the correct answer will still be found to a very high probability.

\vskip 24pt
\noindent{\twelvebf Conclusion}
\vskip \preskip

By allowing the Grover search algorithm to be called as a subroutine
in conjunction with a sort on blocks inside of a larger Grover search,
the block insider of $L_1$ containing the putative match can be
identified.  At the end, this block then will be classically sorted,
at cost $\bigo{N^{1/2}\log N}$ and then $L_2$ will be Grover searched
for the match with an element of this block, again with cost
$\bigo{N^{1/2}\log N}$.  The whole algorithm then takes time
$\bigo{N^{3/4}\log N}$.  

The memory requirement for this whole algorithm is $N^{1/2}$, and this
has to be quantum memory since it is used in superposition for the
inner subroutine call.  For large values of $N$, this alone could be
an obstacle, regardless of how many steps the whole algorithm must
take and maintain quantum coherence.

It is not known whether $\bigo{N^{3/4}\log N}$ is a lower bound on the
matching problem for a quantum computer, but it pretty clearly beats
the classical lower bound which has to be at least $N$, since the two
lists themselves are that long.  It is not known whether there are
faster quantum sorting algorithms that might be employed, but
it is known that faster than classical algorithms exist for insertion
into a sorted list. The algorithm in [FGGS] is better than the best
such classical algorithm, but only by a constant, and it is known from
[A] that a constant speedup is the most that can be hoped for.

\vskip \preskip
\noindent{\twelvebf Bibliography}

\vskip \bskip\noindent
[A] Ambainis, A., `` A Better Lower Bound for Quantum Algorithms
Searching an Ordered List'',
quant-ph/9902053

\vskip \bskip\noindent
[BBHT] Boyer, M., Brassard, G., Hoyer, P., Tapp, A., ``Tight bounds on
quantum searching'', {\it Proceedings of the Fourth Workshop on
Physics of Computation}, 1996, pp. 36-43.

\vskip \bskip\noindent
[BHT] Brassard, G., Hoyer, P., Tapp, A., ``Quantum Algorithm for the
Collision Problem'', 
quant-ph/9705002.

\vskip \bskip\noindent
[FGGS] Farhi, E.,  Goldstone, J, Guttman, S., Sipser, M., ``Invariant
Quantum Algorithms for Insertion into an Ordered List'',
quant-ph/9901059

\vskip \bskip\noindent
[Gr] Grover, L. K., ``A fast quantum mechanical algorithm for
database search'', {\it Proceedings of the $28^{\rm th}$ Annual ACM
Symposium on Computing}, 1996, pp. 212-219.

\vskip \bskip\noindent
[K] Knuth, D. E., {\sl The Art of Computer Programming.  Volume 3:
Sorting and Searching}, Addison-Wesly, Reading, Mass., 1975.

\vskip \bskip\noindent
[NR] Press, W. H., Teukolsky, S. A., Vetterling, W. T., Flannery,
B. P., {\sl Numerical Recipes in C: The Art of Scientific Computing,
$2^{nd}$ Edition}, Cambridge University Press, Cambridge, England, 1992.

\vskip \bskip\noindent
[S] Sedgewick, R., {\sl Algorithms}, Addison-Wesly, Reading, Mass.,
1983. 

\vskip \bskip\noindent
[Z] Zalka, C., ``Grover's Quantum Searching Algorithm is Optimal'',
quant-ph/9711070

\end